\documentclass[12pt]{iopart}
\newcommand{\ox}{\omega_{\rm x}}
\newcommand{\rx}{\rho_{\rm x}}

\newcommand{\be}{\begin{equation}}
\newcommand{\ee}{\end{equation}}

\newcommand{\rvac}{\rho_{\rm vac}}

\newcommand{\oc}{\omega_{\rm ch}}
\newcommand{\rc}{\rho_{\rm ch}}
\newcommand{\pc}{p_{\rm ch}}
\newcommand{\dd}{\rm d}
\newcommand{\rl}{r_{\Lambda}}

\newcommand{\un}{\textbf{u}}

\newcommand{\lp}{\left(}
\newcommand{\rp}{\right)}
   
\newcommand{\dtx}{\dd^{3}{\rm x}}

\newcommand{\dtr}{\dd^{3}{\rm r}}

\newcommand{\maw}{\mathcal{W}}
\newcommand{\mak}{\mathcal{K}}

\newcommand{\mai}{\mathcal{I}}

\usepackage{graphicx}
\begin{document}
\title{From Global to Local Dynamics:\\ Effects of the Expansion on Astrophysical Structures}
\author{A. Balaguera-Antol\'{\i}nez}
\address{Max Planck Institut f\"ur Extraterrestrische Physik,  Garching, Gliessenbachstrasse 1 D-85748, Garching, Germany}
\ead{a-balagu@uniandes.edu.co}
\author{M. Nowakowski}
\address{Departamento de F\'{\i}sica, Universidad de los Andes,A.A. 4976, Bogot\'a, D.C., Colombia.}
\ead{ mnowakos@uniandes.edu.co}
\begin{abstract} 
We explore the effects of background cosmology on large scale structures with non-spherical symmetry by
using the concept of quasi-equilibrium which allows certain internal properties (e.g. angular velocity)
of the bodies to change with time. In accordance with the discovery of the accelerated phase of the universe we model the cosmological background by two representative models: the $\Lambda$CDM Model and the Chaplygin Gas Model. We compare the effects of the two models on
various properties of large astrophysical objects. Different equations of state are also invoked in
the investigation.
\end{abstract}   
\date{}
\maketitle

%=================================================================================================================================================== 
\section{Introduction}
The interest in the impact of the cosmological expansion on bound astrophysical system can be traced back to the paper
by Einstein and Straus \cite{Einstein}. Over the subsequent years different approaches have been used
to gain some insight into the interplay between cosmology and astrophysical structures \cite{Noerdlinger,Bona,Baker,McVittie,nunes}.
With the advent of the discovery of the accelerated Universe \cite{perl}, it seems timely to pick up the topic once again and contrast
the effects of
different models which can explain the current accelerated phase. Indeed, in \cite{Nesseris} such an examination was
performed for phantom \cite{phantom} and quintessence cosmologies \cite{quin}. In the present article we will use
the same approach as \cite{Noerdlinger} and \cite{Nesseris} and apply it to models with a positive cosmological constant 
and Chaplygin gas models. These two models will serve us as representatives for
cosmological models which can explain the current stage of acceleration. 
In addition we will also invoke a general model with variable equation of state of the form
$\omega_{\rm x}=p_{\rm x}/\rho_{\rm x}$ (DECDM). The approach mentioned above consists in generalizing 
the Newtonian limit by allowing terms which are specific to the background cosmology. 
This is not limited to the inclusion of the cosmological constant.  
Notice that the effects of the background cosmology on bound systems are now in general time dependent. This is, however not
in contradiction with the fact that we rightly consider most of the astrophysical structures to be in gravitational
equilibrium. The time dependent background cosmology will not tear the system apart, unless its density is diluted, but change only, 
over cosmological
time, the internal properties of the system (like inner velocities or angular velocity etc). As a mathematical tool we will make use
of the virial theorem (\cite{Cardoso} for different methods) 
by allowing certain internal properties to be time dependent. The virial theorem can be easily derived
even in the presence of the cosmological background as the latter enters only the Newtonian limit (i.e. the Poisson
equation for the gravitational potential which includes now a time dependent part).  

As shown in \cite{bala3} Dark Energy can affect certain astrophysical static properties which have to do with
either the virialization of the system \cite{mota1} or the motion of test bodies \cite{Noerdlinger}. 
For instance, the cosmological constant sets itself scales of distance ($\rl=\Lambda^{-1/2}$)
\footnote{We work in the so-called geometrized units, i.e. $G_N=c=1$}, time and mass, which are of the same order 
of magnitude as the radius of the observed universe, the age of the universe and the total mass of the universe (the so called coincidence problem). 
At the first glance it looks hopeless to expect any effects from $\Lambda$ at astrophysical scales. This is so indeed as long as no other scale
enters the theory. If the latter appears in the theory a combination of the large cosmological and the small internal scale can yield
values of astrophysical relevance (e.g. $(r_s\rl)^{1/3}$ where $r_s$ is the Schwarzschild radius) \cite{bala1}. Furthermore, general conditions for detecting the effects of variations of 'constants' represented by scalar fields  (just like the cosmological constant and quintessence models) at scales ranging from terrestial scales have been well established in \cite{barrow1, barrow2,barrow3}.
Another enhancement mechanism of background cosmology emerges when we are dealing with non-spherically symmetric objects. In such case the ratio of two
length scales to some positive power (bigger than one) goes often hand in hand with $\Lambda$. For this reason, in examining time
dependent effects, we will mostly employ ellipsoidal configurations.
For general consideration of equilibrium in the spherical case see \cite{Boehmer1} and \cite{Boehmer2}, the quasi spherical case
has been discussed in \cite{Debnath}.

In this paper, we will entirely concentrate on effects of Dark Energy followed over cosmological times, past and future.
We do not expect these effects to be large, however, we think that it is of some importance to probe into these matter,
especially in view of the accelerated universe.
In \cite{bala5} we restricted ourselves to effects over small time scales. In this sense these two
articles complement each other. 

%================================================================================================================================
\section{Local dynamics with background cosmology}
In standard cosmology, the universe is described as an ideal fluid which implies the properties of homogeneity and isotropy. Such properties are mathematically represented in Friedman-Robertson-Walker line element, which is written for a flat universe as $\dd s^{2}=-\dd t^{2}+R^{2}(t)\lp\dd r^{2}+r^{2}\dd \Omega^{2}\rp$. Einstein field equations and the FRW line element yields the acceleration equation and the well known Friedman equation, written respectively as
\begin{equation}\label{fri1}
\left[\frac{\dot{R}(t)}{R(t)}\right]^{2}=H(t)^{2}=\frac{8}{3}\pi\rho_{\rm b}(t),\hspace{1cm}
\frac{\ddot{R}(t)}{R(t)}=-\frac{4}{3}\pi \left[\rho_{\rm b}(t)+3p_{\rm b}(t)\right],
\end{equation}
In conventional cosmology, the total energy density $\rho_{\rm b}$ 
is a contribution of a radiation, cold dark matter and dark energy. By neglecting interaction among these components, each one evolves through the conservation equation $\dot{\rho}=-3H(\rho+p)$, so that radiation scales as $\rho_{\rm rad}=\rho_{\rm rad}(R=1) R^{-4}$, cold dark matter scales as $\rho_{\rm cdm}=\rho_{\rm cdm}(R=1) R^{-3}$, while the Dark Energy component associated with the cosmological constant is $\rvac=$ constant.  The major contribution in connection to the cosmological constant is approximately $70\%$ of total energy density \cite{perl}. 
%---------------------------------------------------------------------------------------------------------------------------------
\subsection{Dark energy}
As pointed before, almost the $70\%$ of the content of energy density in the unverse is ruled by a dark energy component, and astronomical observations imply that at the present time this dark energy component might be represented by the cosmological constant. That is, at the present time, the equation of state that displays the best fit with astronomical observation is simply $p_{\rm vac}=-\rvac$, i.e, $\omega=-1$. Nevertheless, several models has been proposed in order to reproduce the present value of dark energy but maintaining a time dependent energy density. One generalizes the equation of state for the dark energy by $p_{\rm x}(R)=\ox(R) \rx(R)$. Dark energy then scales as $\rx(R)=\rx(R=1)a^{-f(R)}$, where the function $f(R)$ is defined as
\begin{equation}
f(R) \equiv \frac{3}{\ln R}\int_{1}^{R}\frac{\omega_{\rm x}(R')+1}{R'}\dd R', 
\end{equation}
The model is  constrained by requiring that at the present time the dark energy density reproduces the vacuum energy density associated with the cosmological constant, that is, $\rho_{\rm x}(R=1)=\rvac$. As pointed before, the case $\omega_{\rm x}=-1$ 
corresponds to the cosmological constant $\rho_{\rm x}=\rvac=\Lambda/8\pi$, while for $0>\omega_{\rm x}>-1$ one stands in the Dark Energy realms. Models with 
$\omega_{\rm x}<-1$ lead to future singularities in the so called \emph{Phantom regime} \cite{phantom}. The description of a dynamical dark energy is compatible with the description of inflationary models. \\
In the DECDM model, the  Friedmann equation and the acceleration equation are written as
\be
\lp\frac{\dot{R}(t)}{R(t)}\rp^{2} =H_{0}^{2}\Omega_{\rm vac}h_{1}(R)\hspace{1cm}\frac{\ddot{R}(t)}{R(t)}=-H_{0}^{2}\Omega_{\rm vac}h_{2}(R)=-\frac{8}{3}\pi \rvac h_{2}(R).
\ee
The functions $h_{1,2}(R)$ are given as 
\begin{eqnarray}\label{fri}
h_{1}(R)&\equiv& \frac{\Omega_{\rm cdm}}{\Omega_{\rm vac}}R^{-3}+R^{-f(R)},\\ \nonumber
h_{2}(R)&\equiv &\frac{1}{2}
\left[\frac{\Omega_{\rm cdm}}{\Omega_{\rm vac}}R^{-3}+R^{-f(z)}\lp1+3\omega_{\rm x}(R)\rp\right],
\end{eqnarray}

\subsection{Chaplygin Gas}
\begin{figure}[t]
\begin{center}
\includegraphics[angle=270,width=12cm]{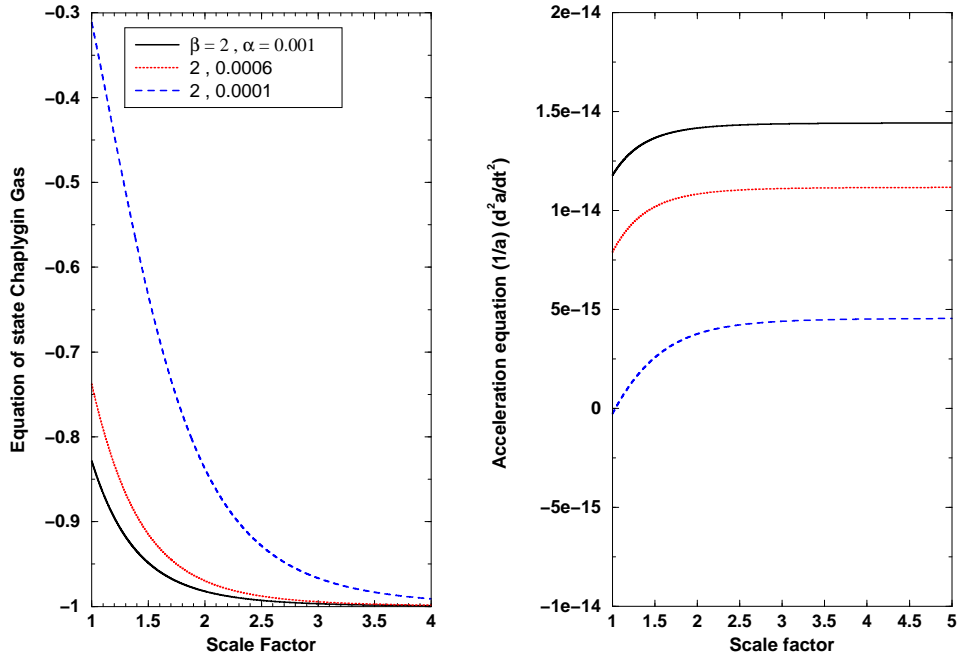}
\end{center}
\caption[]{\footnotesize{Left: Equation of state for Chaplygin gas $\oc=\pc/\rc$ for different Chaplygin parameters. Right: Acceleration equation as a function of the (future) scale factor for different Chaplygin parameters.}}
\label{w}
\end{figure}

The main stages the universe has passed through in the standard cosmology scenario can be reproduced by the introduction of the equation of state of an exotic ideal relativistic gas written as
\begin{equation}
\label{ch1} 
\pc=\kappa_{1} \rc-\frac{\kappa_{2}}{\rc^{\gamma}},
\end{equation}
The relevance of this equation of state lies in three different facts: a) it violates the strong energy condition, which is necessary to obtain an accelerated phase at the present time b) it generates a well definite speed of sound, which is relevant for the process of structure formation and c) it unifies the early radiation or dark-matter behavior with the late dark energy dominance. The so-called pure Chaplygin gas is obtained with $\kappa_{1}=0$ and $\gamma=1$.\\
The time evolution of the Chaplygin Gas-energy density is given from the integration of mass-energy conservation equation as
\begin{equation}
\label{ch2}
\rc(R)=\left[A+BR^{-n\beta}\right]^{\frac{1}{\beta}},\hspace{0.8cm}
\end{equation}
with $n=3(1+\kappa_{1})$, $\beta=\gamma+1$ and where $A$ and $B$ are integration constants given as
\begin{equation}
B=\rc(R=1)^{\beta}-A,\hspace{1.cm}A=\frac{3}{n}\kappa_{2}\equiv \frac{3}{n}\alpha_{n}H_{0}^{2\beta},
\end{equation}
with $\rc(R=1)=\rho_{\rm crit}$ as the Chaplygin Gas energy density at the present time (only for a flat universe). In order to write the integration constants as we have done, we have re-parametrized the Chaplygin Gas equation of state with $\kappa_{2}=\alpha_{n}H_{0}^{2\beta}$ such that a finite age and a present accelerated phase for the Chaplygin gas universe is reached for 
\be \alpha_{n}< \frac{n }{3}\lp\frac{3}{8\pi}\rp^{\beta}.
\ee
This bound can be derived after integration of the Friedmann equation 
\be
H(R)^{2}=\frac{\dot{R}^{2}}{R^{2}}=\frac{8}{3}\pi \left[A+BR^{-n\beta}\right]^{\frac{1}{\beta}}.
\ee
in order to determine the age of the universe. The age of the universe is given in terms of hypergeometric function as
\begin{eqnarray}
T=&&\frac{1}{qH_{0}}\left[\lp\frac{3}{8\pi}\rp^{\beta}-\frac{3}{n}\alpha_{n}\right]^{-\frac{1}{2\beta}}\sqrt{\frac{3}{2\pi}} \\ \nonumber
&&\times 
\,_{2}F_{1}\left[\frac{q}{2n\beta},\frac{1}{2\beta},\frac{2n\beta +n+4}{2n\beta},-\frac{3}{n}\alpha_{n}\left[\lp\frac{3}{8\pi}\rp^{\beta}-\frac{3}{n}\alpha_{n}\right]^{-1}\right],
\end{eqnarray}
with $q\equiv 3\omega+7$. For the pure Chaplygin gas one then obtains a finite age and the observed acceleration at the present time only if the parameter $\alpha$ is such that $\alpha_{3}<10^{-2}$, which in turn implies a bound on the Chaplygin parameter $\kappa_{2}<\rho_{\rm crit}^{2}$. This warranties that the equation of state is $\oc(\rm today)>-1$.\\
The solution given in Eq.(\ref{ch2}) allows us to describe the behavior of Chaplygin gas at different ages. For early times the $\kappa_{1}$-term dominates and Chaplygin Gas behaves as $\rc= B^{1/\beta}R^{-n}$. This scales as radiation for $n=4$ ($\kappa_{1}=1/3$). Hence we could in principle relate the integration constant $B$ with cosmological parameters of the concordance model as $B=\rho^{\beta}_{\rm rad}(R=1)$.  For $\kappa_{1}=0$, the model displays a matter dominated epoch at early times with $\rho_{\rm ch}\approx B^{1/\beta}R^{-3}$. In this case one can write the integration constant $B=\rho^{\beta}_{\rm mat}(R=1)$ and also the constant $A$ as $A=\rho_{\rm crit}^{\beta}(1-\Omega_{\rm mat}^{\beta})$. Finally for large times, Chaplygin gas acquires a vacuum-like behavior with 
\be
\rc(t\to\infty)=A^{1/\beta}=\lp \frac{3\alpha_{n}}{n}\rp^{1/\beta}H_{0}^{2}.
\ee
This would lead to an effective vacuum energy density $\rvac^{\rm eff}$ which can be associated to the \emph{current} vacuum energy density $\rvac$ via 
\be
\rvac^{\rm eff}=\rvac\left[\Omega_{\rm vac}\lp1+\frac{B}{A}\rp^{1/\beta}\right]^{-1}
%\frac{\Lambda^{\rm eff}}{\Lambda}=\frac{\rc(a\to \infty)}{\rc(a=0)}\Omega_{\rm vac}^{-1}.
\ee
With $\rvac^{\rm eff}>\rvac$, the effects of this effective vacuum energy density may be relevant in the Newtonian limit when we explore the equilibrium conditions of large scale structures.\\
Let us concentrate on the case $\kappa_{1}=0$, which does not reproduce an early radiation dominated era but still unifies an early dark matter dominance and a vacuum dominated era at large scale factors. The acceleration equation for this model reads as
\begin{eqnarray}
\label{chacc}
\frac{\ddot{R}(t)}{R(t)}&=&-\frac{4}{3}\pi \rc(R)\eta_{\rm ch}(R)\\ \nonumber
&=&-\frac{4}{3}\pi \left[A+BR(t)^{-3\beta}\right]^{\frac{1}{\beta}}\left[1-\frac{3\kappa_{2}}{A+BR(t)^{-3\beta}}\right].
\end{eqnarray}
where $\eta_{\rm ch}\equiv 1+3\oc$ and 
\be
\oc=\oc(R)\equiv \frac{\pc}{\rc}=-\kappa_{2}\rc(R)^{-\beta}
\ee
is the time dependent equation of state. In Fig. \ref{w} we show the behavior of the equation of state $\oc$ and the acceleration equation for different values of Chaplygin parameters.

%============================================================================================================
\subsection{Newtonian description}
The effects of the background can be explored through the Newtonian limit of field equations \cite{nowakowski2}, from which one can derive a modified Poisson's equation \cite{Noerdlinger}
\be\label{pois}
\nabla^{2}\Phi=4\pi \delta \rho-3\lp\frac{\ddot{R}}{R}\rp,
\ee 
where $\delta\rho$ is the overdensity that gives rise to 
the ``gravitational contribution'' of the potential. The total energy density within the clustered configuration $\rho$ is a contribution of the background $\rho_{\rm b}$ and the collapsed fraction $\delta \rho$. The background energy density depends on the model we are interested in:
\be\label{rhoo}
\rho_{\rm b}=\rho(R)-\delta \rho=  
\cases{\rho_{\rm cdm}(R) & DECDM \\ \rc(R) & Chaplygin}
\ee
Note that the last expression implies that in the DECDM model one has assumed that only the cold dark matter component is what collapses and forms vitalized structures, while in Chaplygin gas the perturbation is done on the total energy density.  The solution for the potential can be simplified as 
\begin{equation}
\Phi(r,R)=\Phi_{\rm grav}(r)-\lp\frac{\ddot{R}}{R}\rp r^{2},\hspace{1cm}\Phi_{\rm grav}(r)=-\int\frac{\delta \rho(r')}{|\textbf{r}-\textbf{r}'|}\dtr',
\end{equation} 
where one has neglected the terms associated to boundary conditions (the potential cannot be zero at infinity) \cite{nowakowski2}. The last term of Eq.(\ref{pois}) reduces to $\rvac$ for $\omega_{\rm x}=-1$ and negligible contribution from the cold dark matter component, that is, for low red-shifts.  This limit represents the Newton-Hooke space time, which aside from the gravitational interaction has an external force of the Hooke form $\sim \Lambda r^{2}$.\\
Gravitational equilibrium is represented through Euler equation $\rho \dot{\un}=-\nabla p-\rho \nabla \Phi$, 
where $\rho$ is given by Eq.(\ref{rhoo}). The next generation of equilibrium equations comes from taking moments on Euler equation and derive the second order virial equation \cite{bala3,nowakowski2,nowakowski1}
\begin{equation}
\label{virialeq}
\frac{\dd ^{2}\mai_{ik}}{\dd t^{2}}=2\mak_{ik}+\maw^{\rm grav}_{ik}+3\Pi_{ik}+\lp \frac{\ddot{R}}{R} \rp \mai_{ik},
\end{equation} 
where $\maw^{\rm grav}_{ik}$ is the gravitational potential energy tensor, whose trace corresponds to the gravitational potential energy, $\mak_{ik}$ is the kinetic energy tensor, $\mai_{ik}$ is the moment of inertia tensor and $\Pi_{ik}$ is the dispersion tensor. The trace of these quantities lead to the well known forms for gravitational potential energy:
\begin{equation}
\label{maw}
\maw^{\rm grav}=-\int_{V}\rho r_{i} \partial_{i} \Phi_{\rm grav}\dtr=\frac{1}{2}\int_{V}\rho\Phi_{\rm grav} \dtr,
\end{equation}
together with $\mai\equiv \int_{V} \rho r^{2}\dtr$ and $\Pi\equiv \int _{V}p\dtr$. As usual, the set of equilibrium equations is closed with the equation for mass conservation, energy conservation and an 
equation of state $p=p(\delta \rho,s)$. \\
%The second derivative of the moment of inertia tensor can be written in this scheme as
%\be\label{sder}
%\frac{\dd^{2} \mai_{ik}}{\dd t^{2}}=R\lp\frac{\ddot{R}}{R}\rp\frac{\dd \mai_{ik}}{\dd R}+R^{2}\lp\frac{\dot{R}}{R}\rp^{2}\frac{\dd^{2} \mai_{ik}}{\dd R^{2}}.
%\ee
The tensor virial equation can be thought as a differential equation for the moment of inertia, which in turns is converted to a differential equation for the parameters determining the geometrical properties of the configuration. In order to explore the contribution from the background, let us write the virial equation (\ref{virialeq}) as
\begin{eqnarray}
\label{tvir}
\frac{\dd ^{2}\mai_{ik}}{\dd R^{2}}=\frac{2}{R^{2}H^{2}(R)}\left[
2T_{ik}+\maw^{\rm grav}_{ik}+\lp\frac{\ddot{R}}{R}\rp \lp\mai_{ik}-\frac{1}{2}\frac{\dd \mai_{ik}}{\dd t}\rp+\Pi_{ik}\right].
\end{eqnarray} 
%===================================================================================================================================================

\section{Dynamical equilibrium}
\noindent If we assume equilibrium via $\ddot{\mai}\approx 0$, we obtain the virial theorem
\begin{equation}
\label{virial}
|\maw^{\rm grav}|=2\mak+\lp\frac{\ddot{R}}{R}\rp\mai,\hspace{1cm}\mak=T+\frac{3}{2}\Pi.
\end{equation} 
Strictly speaking, a formal equilibrium configuration is never reached since the energy-like terms in (\ref{virialeq}) are time dependent. 
Nevertheless, we can use equation (\ref{virial}) by assuming that as long as $R$ evolves in time, the configuration evolves through successive states of equilibrium.\\
The expressions derived in the last section, specially (\ref{pois}) and (\ref{virial}) can be used for testing dark energy models on 
configurations in equilibrium. In this section, we will concentrate on possible effects of the two cosmological models described above on cosmological structures with non-spherical symmetry, such as low density galaxies and galactic clusters.\\
For ellipsoidal homogeneous configurations, the gravitational potential tensor and the moment of inertia tensor are written as
\begin{eqnarray}
\label{wellip}
\maw_{ik}^{\rm grav}&=&-\frac{8}{15}\pi^{2}\rho(\rho-\rho_{\rm b})a_{1}a_{2}a_{3}a_{i}^{2}A_{i}\delta_{ik}=-2\pi\delta \rho A_{i}\mai_{ik},\nonumber \\
\mai_{ik}&=&\frac{4}{15}\pi\rho a_{1}a_{2}a_{3}a_{i}^{2}\delta_{ik},
\end{eqnarray}
where the quantities $A_{i}$ are functions of the eccentricities of each case: for an oblate configuration we have $a_{1}=a_{2}>a_{3}$,  $e_{\rm oblate}^{2}=1-q_{3}^{2}$, while for prolate $a_{1}=a_{2}<a_{3}$,  $e_{\rm prolate}^{2}=1-q_{3}^{-2}$, with $q_{i}\equiv a_{i}/a_{1}$.  The functions $A_{i}$ are given as \cite{binney}:
\begin{eqnarray}
A_{1}=A_{2}=
\cases{\frac{\sqrt{1-e^{2}}}{e^{2}}\left[\frac{\arcsin e}{e}-\sqrt{1-e^{2}}\right] & Oblate \\ 
\frac{1-e^{2}}{e^{2}}\left[\frac{1}{1-e^{2}}-\frac{1}{2e}\ln \lp\frac{1+e}{1-e}\rp\right] & Prolate}
\end{eqnarray}
and
\begin{equation}
A_{3}=
\cases{
2\frac{\sqrt{1-e^{2}}}{e^{2}}\left[\frac{1}{\sqrt{1-e^{2}}}-\frac{\arcsin e}{e} \right]& Oblate \\ 
2\frac{1-e^{2}}{e^{2}}\left[\frac{1}{2e}\ln \lp\frac{1+e}{1-e}\rp-1\right]& Prolate\\
}
\end{equation}
In the limit $e\to 0$ one has $A_{i}\to 2/3$ for the prolate and oblate configurations.
Using Eqs.(\ref{wellip}), the potential energies are then written for the models of interest as

\begin{eqnarray}\label{real}
\maw^{ik}&=&\maw_{ik}^{\rm grav}+\frac{\ddot{R}}{R}\mai_{ik}\\ \nonumber &=&
\cases{
-\frac{8}{15}\pi^{2}\rho^{2}a_{1}a_{2}a_{3}a_{i}^{2}A_{i}\delta_{ik}-\frac{4}{3}\pi \eta_{\rm x}\rho_{\rm x}\mai_{ik}+2\pi \rho_{\rm cdm}\lp A_{i}-\frac{2}{3}\rp\mai_{ik} \\
-\frac{8}{15}\pi^{2}\rho^{2}a_{1}a_{2}a_{3}a_{i}^{2}A_{i}\delta_{ik}
+2\pi \rc \lp A_{i}-\frac{2}{3}\eta_{\rm ch}\rp\mai_{ik}&\\
}
\end{eqnarray}
for DECDM and Chaplygin Gas respectuvely, with $\eta_{\rm x}=1+3\omega_{\rm x}$ and $\rho_{\rm x}=\rvac (1+z)^{f(z)}$.  Note that the last term on the DECDM model vanishes in two different situations: the first is in the case of spherical symmetry, and when we neglect the contribution from the cold dark matter component with respect to the proper density of the system and the dark energy contribution. This approximation is only valid for very low red-shifts. At high red shifts we approach a matter dominated universe and hence the term proportional to $\rho_{\rm cdm}$ is relevant. 

%=============================================================================
\subsection{Oblate systems}
When considering oblate configurations ($a_{1}=a_{2}>a_{3}$), we assume that the kinetic energy is due to constant rotation along the minor axis. We then may write the kinetic energy tensor as 
\begin{equation} 
\label{ket2}
T_{ik}=\frac{1}{2}\lp\Omega^{2}_{\rm rot}\mai_{ik}-\Omega_{{\rm rot}i}\mai_{kj}\Omega_{{\rm rot}j}\rp, \nonumber
\end{equation}
In order to determine the angular velocity, one may consider a non-zero isotropic dispersion tensor $\Pi_{ik}=\delta_{ik}\Pi$. We then have
\begin{equation}
\label{ns01al}
\Omega_{\rm rot}^{2}=\frac{|\maw_{xx}^{\rm grav}|-|\maw_{zz}^{\rm grav}|}{\mai_{xx}}-
\lp\frac{\ddot{R}}{R}\rp \lp1-\frac{\mai_{zz}}{\mai_{xx}}\rp,
\end{equation}
which follows from eliminating the trace of the dispersion tensor from the tensor virial equations. Using Eq.(\ref{real}), we can explicitly write for the angular velocity
\begin{eqnarray}
\label{anr} 
\frac{\Omega_{\rm rot}^{2}}{2\pi \delta \rho}=&&\lp
A_{1}-A_{3}q_{3}^{2}\rp \times \\ \nonumber
&&\lp1+\frac{1-q_{3}^{2}}{2\pi \delta \rho(A_{1}-A_{3}q_{3}^{2})}\times \cases{
H_{0}^{2}\Omega_{\rm vac}h_{2}(R) & DECDM \\
\frac{4}{3}\pi \rc(R)\eta_{\rm ch}(R)& Chaplygin}\rp
\end{eqnarray}
In the Newton-Hooke space time with $\omega_{\rm x}=-1$ ($h_{2} \to -1$) this reduces to 
\begin{equation}
\label{anr2}
\frac{\Omega_{\rm rot}^{2}}{2\pi \delta \rho}\approx A_{1}-A_{3}q_{3}^{2}-\frac{4}{3}\lp\frac{\rvac}{\delta \rho}\rp(1-q_{3}^{2}).
\end{equation}
which in turn reproduces the Maclaurin formula for $\rvac=0$.
%==================================================================================================================
\subsection{Prolate systems}
\begin{figure}[t]
\begin{center}
\includegraphics[angle=270,width=12cm]{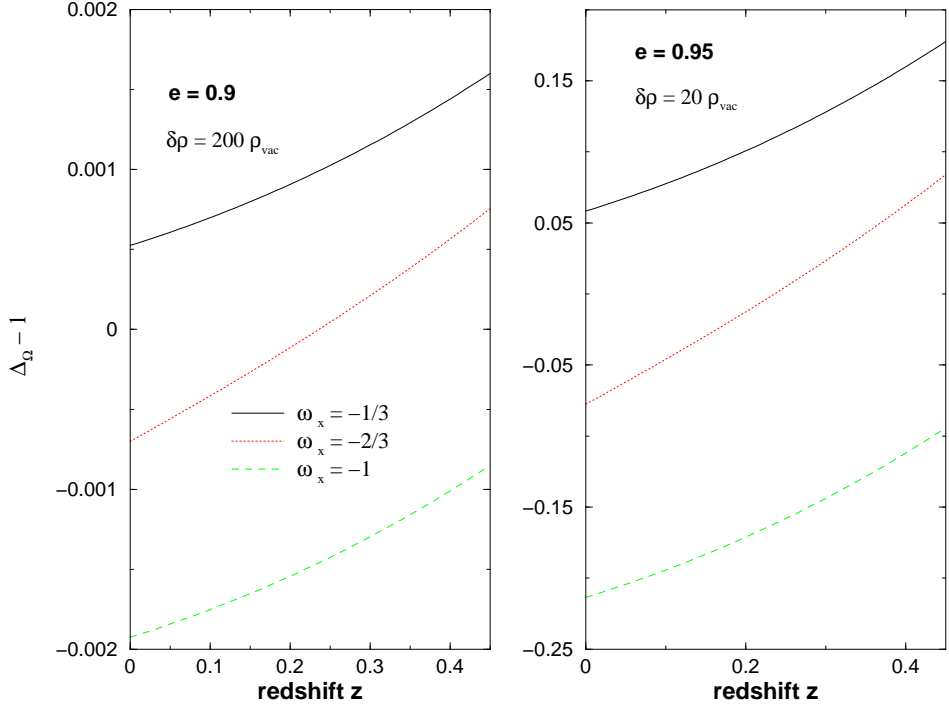}
\end{center}
\caption[]{\footnotesize{Equation (\ref{D}) for $\Delta_{\Omega}-1$ in terms of the red-shift (up to $a\sim 0.7$), for different densities and eccentricities in a DECDM  model for typical values of equation of state. \cite{bala4}}}
\label{v2}
\end{figure}

For prolate configurations ($a_{1}=a_{2}<a_{3}$), we solve for the velocity dispersion of main components rather than for an angular velocity (rotation of prolate configurations, although rare, has been observed and some properties of this situation has been explored in \cite{bala5}). Hence we use the virial theorem (\ref{virial}) with a the kinetic energy given by
\begin{equation}
\mak=\frac{1}{2}\int_{V}\rho \langle v^{2}\rangle \, \dtr=\frac{2}{3}\pi \rho a_{1}^{2}a_{3} \langle v^{2}\rangle.
\end{equation}
In analogy with Eq.(\ref{anr}), the virial theorem then implies for the velocity dispersion 
\begin{equation}
\frac{\langle v^{2}\rangle}{2\pi \delta \rho}= \frac{1}{5}a_{1}^{2}\lp 2A_{1}+q^{2}_{3}A_{3} \rp
\lp 1+\frac{2+q_{3}^{2}}{2\pi \delta \rho(2A_{1}+q^{2}_{3}A_{3})}  \times 
\cases{
H_{0}^{2}\Omega_{\rm vac}h_{2}(R) \\
\frac{4}{3}\pi \rc(R)\eta_{\rm ch}(R)}\rp
\end{equation}
for the DECDM and Chaplygin Gas model respectively. As in (\ref{anr}), one obtains a scale factor (or red-shift) dependence of the velocity dispersion through the contribution of the cosmological background. In the Newton-Hooke space-time one reduces to
\begin{equation}\label{veld2}
\frac{\langle v^{2}\rangle }{2\pi \delta \rho} \approx \frac{1}{5}a_{1}^{2} (2A_{1}+q_{3}^{2}A_{3})\left[1-\frac{4}{3} \frac{\rvac}{\delta \rho}\lp\frac{2+q_{3}^{2}}{2A_{1}+q_{3}^{2}A_{3}}\rp\right],
\end{equation}
which in turn reduces for spherical symmetry to the known expression \cite{bala3,wang}
\begin{equation}\label{veld3}
\frac{\langle v^{2}\rangle }{2\pi \delta \rho} \approx \frac{1}{5}a^{2} \left[1-2\frac{\rvac}{\delta \rho}\right].
\end{equation}
%\begin{figure}[t]
%\begin{center}
%\includegraphics[angle=270,width=7cm]{g.eps}
%\end{center}
%\caption[]{\footnotesize{Geometrical factor $\mathcal{G}$.}}
%\label{v1}
%\end{figure}
\noindent The structure of the resulting expressions for the angular velocity and the velocity dispersion can be written in a similar way. Let $\Delta$ measures the ratio of each velocities with respect to the velocities when there is not background contribution. We then may write
\begin{equation}
\label{D}
\Delta_{\Omega,v}
\equiv 
\cases{
\frac{\Omega^{2}_{\rm rot}}{\Omega^{2}_{\rm rot}(\rho_{b}=0)}&\\
\frac{\langle v^{2}\rangle }{\langle v^{2}\rangle(\rho_{b}=0)}&
}
=1-\mathcal{G}_{\Omega,v}\lp\frac{1}{2\pi \delta \rho}\rp\frac{\ddot{R}}{R},
\end{equation}
where the geometrical factor $\mathcal{G}$ is written for each case as
\begin{equation}
\mathcal{G}\equiv
\cases{
\mathcal{G}_{\Omega}=\frac{1-q_{3}^{2}}{A_{1}-q^{2}_{3}A_{3}} & Oblate, rotational velocity\\
\mathcal{G}_{v}=\frac{2+q_{3}^{2}}{2A_{1}+q_{3}^{2}A_{3}}& Prolate, velocity dispersion}
\end{equation}
\begin{figure}[t]
\begin{center}
\includegraphics[angle=270,width=12cm]{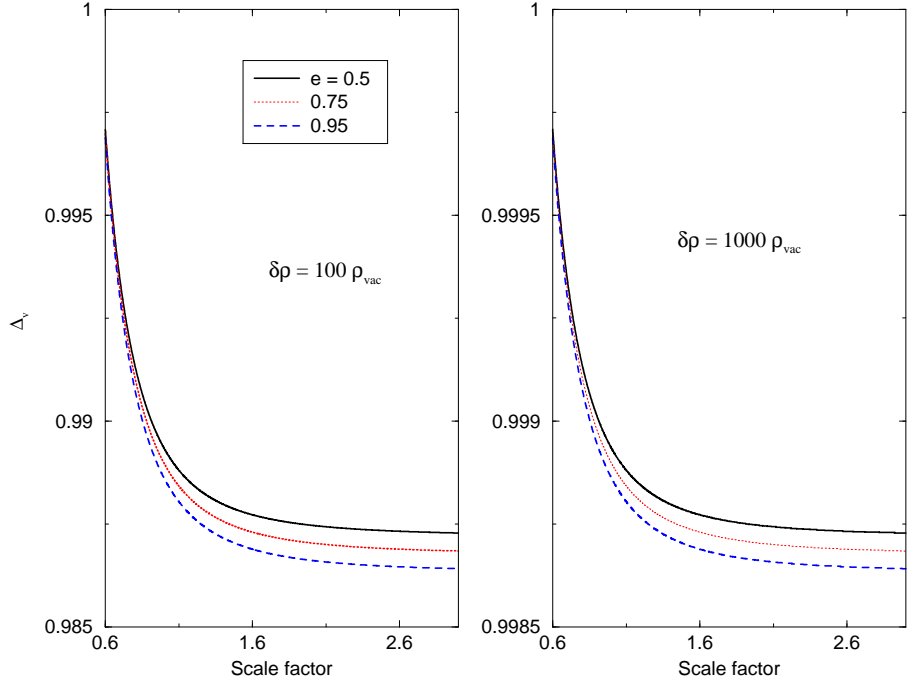}
\end{center}
\caption[]{\footnotesize{Equation (\ref{D}) for $\Delta_{v}$, for different densities and eccentricities in a $\Lambda$CDM model.}}
\label{v1}
\end{figure}
\begin{figure}[t]
\begin{center}
\includegraphics[angle=270,width=12cm]{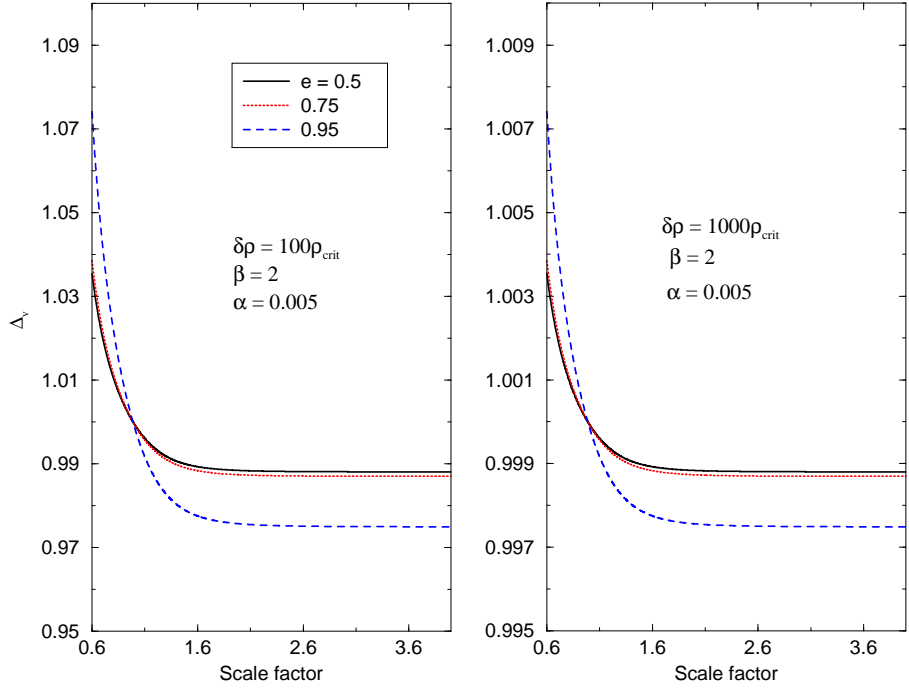}
\end{center}
\caption[]{\footnotesize{Equation (\ref{D}) for $\Delta_{v}$, for different densities and eccentricities in a Chaplygin Gas model for typical values of Chaplygin parameters.}}
\label{v1a}
\end{figure}
Returning to the oblate case, in a generalized Newton-Hooke space-time (i.e, with $-1<\omega<0$ ), one observes that the Dark Energy component decreases the angular velocity (or the velocity dispersion in the prolate case) in order for the system to maintain equilibrium  (since $1+3\omega_{\rm x}<0$). At the present time ($R=1$, $z=0$) we get $\Delta_{\Omega}\sim 1-0.8 g(e)(\rvac/\delta \rho)$, where 
\begin{equation}
\label{H}
g(e)\equiv\frac{4}{3}e^{5}\left[(1-e^{2})^{1/2}(3-2e^{2})\arcsin e-3e(1-e^{2})\right]^{-1}.
\end{equation}
An extreme situation is reached in the limit $\Delta_{\Omega}\to 0$, which implies $\Omega^{2}\to 0$. 
For $e\sim 0.8$ this would require $\zeta\equiv 2\rvac/\delta\rho\sim 0.5$, which is a very diluted configuration. 
For higher eccentricities (say $e\sim 0.97$) one would need a system with $\zeta\sim 0.2$ which is still a rather low value. 
In the $\Lambda$CDM model $\Omega=0$ is reached at a red-shift given by \cite{bala4} $z_{c}=\left[(2\Omega_{\rm vac}(\zeta g(e)-2))/(\zeta g(e)\Omega_{\rm cdm})\right]^{1/3}-1$.
For realistic examples as   $\delta \rho\sim 200 \rho_{\rm cdm}$ the value $z_{c}$ could be reached for $e$ very close to $1$. That is, 
the value $z_{c}$ requires extreme flat objects. \\
In figures \ref{v1} and \ref{v1a} we show the behavior of the ratio $\Delta_{v}$ for different values of eccentricities and proper densities in the DECDM case with $\ox=-1$ and in the Chaplygin Gas model, respectively. The larger effects appears for low densities and large eccentricities, as expected. However, the fractional change is very low. For $\delta \rho=10^{2}\rvac$, the larger effect occurs for scale factors $\geq 2.6$ with a change of $1\%$.\\
In Fig. \ref{v2} we show the behavior of $\Delta_{\Omega}-1$ as a function of the red-shift for different densities and $e=0.9,0.95$ 
for three different values of the equation of state for Dark Energy. 
It is clear that the effects associated with a cosmological constant are 
stronger than the ones associated with other Dark Energy models, 
but in general those effects are small for realistic values of $\zeta$ as used in the figure. We would have to measure the angular 
velocity at different red-shifts very exactly to see an effect over a range of $z$.
However, the difference between the models is more significant.  

%====================================================================================================================================

\section{Dynamical evolution} 
In this section we want to explore the virial equation with the cosmological background 
without insisting on the equilibrium condition (dynamical equilibrium). Even though we will not make
extensive use of these new equations, we want to demonstrate that the equilibrium condition is not
a necessary ingredient while studying the effects of different background cosmologies.
The results are differential 
equation for the geometrical parameters of a given configuration. Our starting point is equation (\ref{tvir}). 
In the first approach, we will concentrate on pressurless systems with homogeneous density. 
The problem is the reduced to determine $a_{1}=a_{1}(R)$ and $a_{3}=a_{3}(R)$. As pointed out before, 
Eq.(\ref{tvir}) represents a set of two non-linear coupled second order equations for $i=x,z$
\be
\label{tvir2}
\frac{\dd ^{2}\mai_{ii}}{\dd R^{2}}=\frac{2}{R^{2}H^{2}(R)}\left[
2T_{ii}+\maw^{\rm grav}_{ii}+\lp\frac{\ddot{R}}{R}\rp \lp\mai_{ii}-\frac{1}{2}\frac{\dd \mai_{ii}}{\dd t}\rp\right].
\ee
where we have used Eq.(\ref{real}). The derivatives of the moment of inertia tensor are given as follows
\begin{eqnarray}
\label{tvir3}
\frac{\dd \mai_{ii}}{\dd R}&=&\frac{8}{15}\pi \rho a_{1}^{2}a_{3}a_{i}\frac{\dd a_{i}}{\dd R}=2a_{i}^{-1}\frac{\dd a_{i}}{\dd R}\mai_{ii},\\ \nonumber
\frac{\dd^{2} \mai_{ii}}{\dd R^{2}}&=&\frac{8}{15}\pi \rho a_{1}^{2}a_{3}a_{i}\left[ \lp a_{i}\frac{\dd a_{i}}{\dd R}\rp^{2}+ a_{i}\frac{\dd a_{i}}{\dd R}\right]=2a_{i}^{-1} \left[\lp a_{i}\frac{\dd a_{i}}{\dd R}\rp^{2}+ a_{i}\frac{\dd^{2} a_{i}}{\dd R^{2}}\right]\mai_{ii},
\end{eqnarray}
These expressions enclose our approximation: {\it the volume of the system is constant even if the semi-axes may change}, so that one can write the moment of inertia tensor as 
\be
\mai_{ik}=\frac{1}{5}Ma_{i}^{2}\delta_{ik},
\ee
where the mass $M$ is
\begin{equation}
M=M_{\rm b}+\delta M=\frac{4}{3}\pi a_{1}^{2}a_{3}\rho_{\rm b}(R)+\delta M,\hspace{1cm}\delta M=\int_{V}\delta \rho\,\dtr.
\end{equation}

In order to deal with the kinetic energy tensor, we follow \cite{peebook} and transform from proper coordinates to fixed co-moving coordinates via $r_{i}=A_{i\alpha}x_{\alpha}$, where $A_{i\alpha}$ is such that $A_{i\alpha}=a_{i}\delta_{i\alpha}$ and hence the ellipsoid is characterized by the constraint $\delta_{\alpha \beta}x_{\alpha}x^{\beta}=1$. We have
\begin{equation}
T_{ik}=\frac{1}{2}\rho\int_{V}\lp\frac{\dd A_{i\alpha}}{\dd t}\rp\lp\frac{\dd A_{k\beta}}{\dd t}\rp x_{\alpha}x_{\beta}A\dtx,
\end{equation}
where $A={\rm det}(A_{\alpha \beta})=a_{1}a_{2}a_{3}$. Using the definition of the moment of inertia tensor one has
\be
\mai_{ik}=\rho A_{i\alpha}A_{k\beta}A\int x_{\alpha}x_{\beta}\dtx\,, 
\ee
and hence
\be
T_{ik}=\frac{1}{2}A_{i\alpha}^{-1}A_{k\beta}^{-1}\lp\frac{\dd A_{i\alpha}}{\dd t}\rp\lp\frac{\dd A_{k\beta}}{\dd t}\rp\mai_{ik}.
\ee
We then may write
\begin{equation}\label{kin}
T_{ii}=\frac{1}{2}a_{i}^{-2}\lp\frac{\dd a_{i}}{\dd t}\rp^{2}\mai_{ii}=a_{i}^{-2}R^{2}\lp\frac{\dd a_{i}}{\dd R}\rp^{2}\mai_{ii} \times
\cases{
\frac{1}{2}H_{0}^{2}\Omega_{\rm vac}h_{1}(R) \\
\frac{4}{3}\pi \rc(R)}
\end{equation}
for the DECDM and Chaplygin Gas respectively. Equation (\ref{tvir2}) together with (\ref{tvir3}) are the differential equation to be solved. The functions $A_{i}$ depend on the eccentricity of the system and hence they are also function of the semiaxis $A_{i}(z)=A_{i}(a_{1}(R),a_{3}(R))$. The differential equation can then be cast into the following form 
\be\label{deq}
\frac{\dd^{2}a_{i}}{\dd R^{2}}=\frac{1}{R^{2}H^{2}(R)}\left[
\frac{2T_{ii}+\maw_{ii}}{\mai_{ii}}-\lp\frac{\ddot{R}}{R}\rp\frac{1}{a_{i}}\frac{\dd a_{i}}{\dd R}
\right]-a_{i}\lp\frac{\dd a_{i}}{\dd R}\rp^{2}.
\ee

%===============================================================================================================================

\section{Conclusion}
In this article we have examined the effects of the expansion of the Universe on certain
quasi-static properties of large astrophysical bodies. We have done this by invoking a dynamical equilibrium
i.e. the time dependent response of the astrophysical objects to the time dependent cosmological background
is taken into account by allowing some internal properties to become epoch dependent. In spite of the fact
that the expansion of the Universe is accelerated the effects are rather small, at the most few per cents over a
large cosmological time stretch in the case of the Chaplygin gas model. 
However, a qualitative investigation, done in the present paper, seems mandatory to establish the size of the effects.
Different equations
of state can be clearly distinguished theoretically, however, in practice this would require a good knowledge of the
angular velocities at different redshifts. In spite of the smallness of the effects, the results clearly demonstrate
that background cosmology affects, in principle, local properties of astrophysical bodies and furthermore, the details
of these effects are model-dependent. In this paper, we focused on the the dynamical equilibrium.
One can enlarge the concept of the influence of background cosmology by dropping the equilibrium condition and go over to a
fully dynamical scenario. A first step in this direction was done on section 4. Supplementing these equation with
Euler equation for the angular velocity, one has a full set of differential equations determining the time evolution
of the parameters during (and after) virialization. We will come back to this point in future publications.

\section*{References}
\bibliographystyle{unsrt} 

\end{document}